\documentclass[11pt,twoside]{article}

\usepackage{asp2014}

\aspSuppressVolSlug
\resetcounters

\begin{document}

\title{AI Agents for Ground-Based Gamma Astronomy}

\author{
    Dmitriy~Kostunin$^{1}$,
    Vladimir~Sotnikov$^{2}$,
    Sergo~Golovachev$^{3}$, and
    Alexandre~Strube$^{4}$
}
\affil{$^1$Deutsches Elektronen-Synchrotron DESY, 15738 Zeuthen, Germany; \email{dmitriy.kostunin@desy.de}}
\affil{$^2$JetBrains Limited, 8046 Paphos, Cyprus}
\affil{$^3$JetBrains GmbH, 80639 M\"unchen, Germany}
\affil{$^4$J\"ulich Supercomputing Centre, 52425 J\"ulich, Germany}

\paperauthor{Dmitriy~Kostunin}{dmitriy.kostunin@desy.de}{0000-0002-0487-0076}{Deutsches Elektronen-Synchrotron DESY}{}{Zeuthen}{}{15738}{Germany}
\paperauthor{Vladimir~Sotnikov}{vladimir.sotnikov@jetbrains.com}{}{JetBrains Limited}{}{Paphos}{}{8046}{Cyprus}
\paperauthor{Sergo~Golovachev}{sergey.golovachev@jetbrains.com}{}{JetBrains GmbH}{}{M\"unchen}{}{80639}{Germany}
\paperauthor{Alexandre~Strube}{a.strube@fz-juelich.de}{}{Author3 Institution}{}{J\"ulich}{}{52425}{Germany}



\begin{abstract}
Next-generation instruments for ground-based gamma-ray astronomy are marked by a substantial increase in complexity, featuring dozens of telescopes.
This leap in scale introduces significant challenges in managing system operations and offline data analysis.
Methods, which depend on advanced personnel training and sophisticated software,
become increasingly strained as system complexity grows,
making it more challenging to effectively support users in such a multifaceted environment.
To address these challenges, we propose the development of AI agents based on instruction-finetuned large language models (LLMs).
These agents align with specific documentation and codebases,
understand the environmental context, operate with external APIs,
and communicate with humans in natural language.
Leveraging the advanced capabilities of modern LLMs, which can process and retain vast amounts of information,
these AI agents offer a transformative approach to system management and data analysis by automating complex tasks and providing intelligent assistance.
We present two prototypes that integrate with the Cherenkov Telescope Array Observatory pipelines for operations and offline data analysis.
The first prototype automates data model implementation and maintenance for the Configuration Database of the Array Control and Data Acquisition (ACADA).
The second prototype is an open-access code generation application tailored for data analysis based on the Gammapy framework.
\end{abstract}



\section{Introduction}
\label{sec:introduction}

Large language models (LLMs)~\citep{2023arXiv230318223Z} are systems trained to perform tasks like causal language modeling,
enabling them to address a wide range of challenges.
However, these models often encounter limitations in tasks requiring logic, calculation, or access to external knowledge.
To address these shortcomings, the concept of AI agents has emerged as a transformative approach.

LLMs can significantly expand their capabilities when connected to tools that provide access to the real world.
This might include the ability to interact with search engines, calculators, or specialized programs, enabling them to perform tasks beyond their standalone functionality.
In other words, granting LLMs a degree of agency allows them to overcome many of their inherent limitations and handle more complex problems.

AI agents are systems that provide this agency.
They use the outputs of an LLM to control workflows, enabling the integration of external tools that support decision-making and task execution.
By combining LLMs with these external functionalities, AI agents effectively bridge the gap between language understanding and actionable outcomes.

Such an agent may be highly useful for astronomy as a ``copilot'' for telescope control,
applied for monitoring, alarm and reporting systems, and used in data quality assurance and analysis.

The exhibition of intelligent behavior is mimicked by chain-of-thought techniques~\citep{2022arXiv220111903W} and reasoning introduced for proprietary \texttt{o1} models by OpenAI.
Learning from experience is simply achieved through validation against field-specific software and data, either real or synthetic.
This is a crucial feature used for the agents described in this paper.
Finally, decision-making can be implemented through function calling triggered by a user or external conditions.

In this way, the interaction of an agent with both user and environment can be depicted as shown in the left pane of Figure~\ref{fig:1}:
the validation pipeline is added to the common route of interaction between an LLM and a user.
The validation implies function calling of the commands written by the LLM within the framework containing sample data;
using the function call's return, the framework evaluates the quality of the command and either repeats the step with command generation and function calling, or proceeds further.

\articlefigure{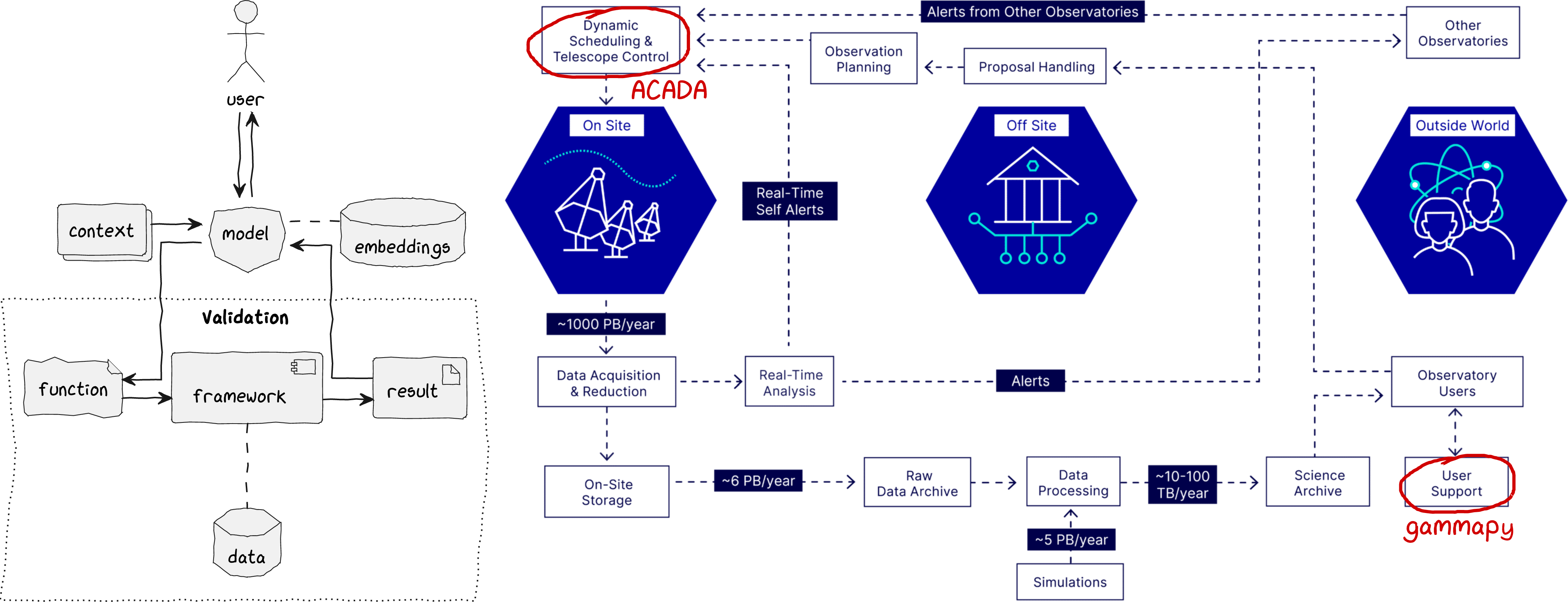}{fig:1}{
	\textit{Left:} schematic representation of the astronomical agent with an abstract depiction of the validation step.
	\textit{Right:} data flow of CTAO (drawing from the official website). We marked with red the parts of the flow addressed in this work.}

\section{Designing agents for astronomy}
\label{sec:designing-agents-for-astronomy}

Modern ground-based observatories comprise sophisticated data pipelines in which data products evolve from raw camera signals to high-level reconstructions for use by external scientists.
In this work, we focus on gamma-ray telescopes, particularly the next-generation Cherenkov Telescope Array Observatory\footnote{\url{https://ctao.org}}.
Its data flow is depicted in the right pane of Figure~\ref{fig:1}.

\subsection{Agent for telescope control}
\label{subsec:agent-for-telescope-control}

One can imagine diverse applications of LLMs for telescope control, e.g., from operational co-piloting to report writing.
As a first exercise, we assess their capabilities for understanding and describing data structures.
Our test uses the framework of a Configuration Database (CDB), a subsystem from Array Control and Data Acquisition~(ACADA) of CTAO~\citep{Oya:2024dww}.
A draft of the structure configuration note for the medium-size telescope~\citep{Bradascio:2023lmr} is provided as context to the \texttt{o1-preview} model in LaTeX format, which was prompted as follows:
\textit{
    Analyze the text of this book in latex. It describes the configuration model of a telescope.
    You need to provide Pydantic classes that describe the configuration model.
    Use as many classes as needed; they can (and probably should) be nested. Here’s the text: <...>
}
As a result, we obtained Pydantic\footnote{\url{https://github.com/pydantic/pydantic}} classes derived from a written description of configuration structures.
To make a shallow analysis of the results, we prompted the \texttt{o1-preview} model to compare these model-generated classes with human-generated JSON schemas:
\textit{
    Compare your models with these JSON schemas and provide a detailed analysis: <...>
}
It is worth noting that we did not invoke any function calling at this moment, i.e., we did not produce JSON schemas from Pydantic classes.
As a result, the model generated a detailed comparison of these two formal descriptions, showing strong agreement between them.
We manually verified the few discrepancies detected by the model; its findings illustrate the efficiency of LLMs in detecting non-conformities with specifications, requirements, documentation, etc.
As a last step in this exercise, we combined the resulting Pydantic models with the core API of ACADA CDB and its tutorial code in the same context and used the following prompt:
\textit{
    Here's a tutorial that demonstrates how you can upload Pydantic models to the telescope’s configuration database: <...>
    Here's the 'cdb.py' from 'acadacdb.core': <...>
    Here are Pydantic models you will need to use: <...>
    Write the code that adds all top-level configurations from the script above to the database (backend).
}
Drawing on the provided tutorial, the model generated a script to maintain CDB entries.
Since only schemas were provided to the model, it included dummy values for the entries based on the context.
Moreover, it reconstructed the naming conventions from the tutorial.
After executing this script, the correct entries appeared in the CDB\@.
Given the successful manual completion of this pipeline, our next obvious step is to automate all aforementioned actions.

\subsection{Agent for data analysis}
\label{subsec:agent-for-data-analysis}

Encouraged by the success of manual code generation for CTAO ACADA,
we proceeded to develop an autonomous agent for code generation following the scheme shown in Figure~\ref{fig:1}.
We chose Gammapy~\citep{gammapy:2023} as the most common cross-observatory analysis tool for high-energy gamma-ray data.
Because Gammapy is an actively developed package, its multiple incompatible versions included in various LLM training sets often lead to invalid code generation.
To address this challenge,
we created an open-access service code-named AstroAgent\footnote{Prototype: \url{https://majestix-vm8.zeuthen.desy.de/}}.
This agent features field-specific prompts and Retrieval-Augmented Generation (RAG) using embeddings based on snippets from Gammapy code, including API parts, tutorials, and documentation.
Based on the initial prompt from the user, the agent attempts to generate valid Python code.
It is worth noting that we take user prompts as is, without additional editing; however, we plan to add a supplemental agent to modify user prompts for code generation optimization.
Currently, we expect relatively precise prompts that assume some familiarity with Gammapy,
e.g.:
\textit{
    The data for gammapy analysis are stored in \$PHOTON\_STORAGE.
    Generate a code which selects available observations of Crab Nebula.
    Using these observations, save a plot with a significance map (sqrt\_ts) based on the RingBackground method.
    Use an exclusion mask for background maker.
}
Validation of the generated code is performed using data in Gammapy format, either synthetic (simulated) or real, e.g.,
science data challenge datasets or public data releases.
For our agent, we use H.E.S.S.\ DL3 public test data release 1~\citep{HESS:2018zix}.
This small dataset allows relatively quick validation. After thorough testing, we managed to fine-tune a GPT-4o model.
Hence, the prototype application is equipped with fine-tuned GPT-4o, \texttt{o1-preview}, and Llama~3.3 70B models deployed at Blablador\footnote{\url{https://helmholtz-blablador.fz-juelich.de/}}.

\section{Conclusion}
\label{sec:conclusion}

It is fair to say that the potential of LLMs in astronomy is still not fully explored.
In our experiments described here, we have seen surprisingly promising results (possibly boosted by relatively low initial expectations).
It has already been demonstrated that LLMs enhance data modeling and code generation, particularly for CTAO tasks at the design or prototyping phase.

Due to the lack or absence of training data in this specialized field,
human expertise is still crucial for designing, verification, and fine-tuning agents.
Another important aspect relates to common machine-learning concerns of reproducibility and interpretability.
We plan to address these aspects by transitioning to open-source LLMs in our future developments.

\acknowledgements
We thank the Cherenkov Telescope Array Observatory\footnote{The authors gratefully acknowledge financial support from the agencies and organizations listed here: \url{https://www.ctao.org/for-scientists/library/acknowledgments/}}, ACADA Collaboration, and CTAO Medium-Size Telescope team for providing materials.
This work made use of data from the H.E.S.S.\ DL3 public test data release 1 analyzed with the Gammapy framework.

\bibliography{C202}

\begin{thebibliography}{}
\expandafter\ifx\csname natexlab\endcsname\relax\def\natexlab#1{#1}\fi
\expandafter\ifx\csname url\endcsname\relax
  \def\url#1{\texttt{#1}}\fi
\expandafter\ifx\csname urlprefix\endcsname\relax\def\urlprefix{URL }\fi
\providecommand{\eprint}[2][]{\url{#2}}

\bibitem[{Abdalla et~al.(2018)}]{HESS:2018zix}
Abdalla, H., et~al. (H.E.S.S.) 2018. \eprint{1810.04516}

\bibitem[{Bradascio et~al.(2023)}]{Bradascio:2023lmr}
Bradascio, F., et~al. (CTA MST Project) 2023, PoS, ICRC2023, 859.
  \eprint{2310.02127}

\bibitem[{{Donath} et~al.(2023)}]{gammapy:2023}
{Donath}, A., et~al. 2023, A\&A, 678, A157.
  \urlprefix\url{https://doi.org/10.1051/0004-6361/202346488}

\bibitem[{Oya et~al.(2024)}]{Oya:2024dww}
Oya, I., et~al. 2024, Proc. SPIE Int. Soc. Opt. Eng., 13101, 131011D

\bibitem[{{Wei} et~al.(2022){Wei}, {Wang}, {Schuurmans}
  et~al.}]{2022arXiv220111903W}
{Wei}, J., {Wang}, X., {Schuurmans}, D., et~al. 2022, arXiv e-prints,
  arXiv:2201.11903. \eprint{2201.11903}

\bibitem[{{Zhao} et~al.(2023){Zhao}, {Zhou}, {Li} et~al.}]{2023arXiv230318223Z}
{Zhao}, W.~X., {Zhou}, K., {Li}, J., et~al. 2023, arXiv e-prints,
  arXiv:2303.18223. \eprint{2303.18223}

\end{thebibliography}

\end{document}